**Prolonged Phase Segregation of Mixed-Halide Perovskite Nanocrystals in the Dark**

Xueying Ma[1,†], Yuhui Ye[1,†], Yang Xiao[1], Shengnan Feng[1], Chunfeng Zhang[1], Keyu Xia[2], Fengrui Hu[2], Min Xiao[1,3,*], and Xiaoyong Wang[1,*]

[1]*National Laboratory of Solid State Microstructures, School of Physics, and Collaborative Innovation Center of Advanced Microstructures, Nanjing University, Nanjing 210093, China*

[2]*College of Engineering and Applied Sciences, Nanjing University, Nanjing 210093, China*

[3]*Department of Physics, University of Arkansas, Fayetteville, Arkansas 72701, USA*

[*]Correspondence to M.X. (mxiao@uark.edu) or X.W. (wxiaoyong@nju.edu.cn)

[†]These authors contributed equally to this work

**A critical issue hindering the potential applications of semiconductor mixed-halide perovskites is the phase segregation effect, wherein localized regions enriched with one type of halide anions would be formed upon continuous photogeneration of the excited-state charge carriers. These unexpected phases are capable of remixing again in the dark under the entropic driving force, the process of which are now being exclusively studied after mixed-halide perovskites have arrived at the final stage of complete phase segregation. Here we show that after the removal of laser excitation from a solid film of mixed-halide perovskite nanocrystals with partial phase segregation, the iodide- and bromide-rich regions can continuously grow in the dark for a prolonged time period of several minutes. We propose that this dark phase segregation is sustained by the local electric fields associated with the surface-trapped charge carriers, whose slow dissipation out of mixed-halide perovskite nanocrystals causes a delayed occurrence of the reversal phase remixing process.**

**Introduction**

Lead mixed-halide perovskites with the chemical formular of $APbBr_xI_{3-x}$ ($0 < x < 3$), where A is an organic or inorganic cation such as $CH_3NH_3^+$ ($MA^+$), $HC(NH_2)_2^+$ ($FA^+$) or Cs+, have attracted intensive research interests very recently owing to their facile synthesis in solution[1], high quantum efficiency of fluorescence[2] and tunable emission across the visible to near-infrared wavelength range[3]. Unfortunately, under continuous photogeneration of the excited-state charge carriers, they would suffer from the phase segregation effect to demix into the bromide- and iodide-rich regions with different bandgap energies[4], thus greatly jeopardizing their stable performances in the optoelectronic devices of solar cells[5,6] light-emitting diodes[7,8] and photodetectors[9,10]. Among the various underlying mechanisms proposed so far for the phase segregation effect of mixed-halide perovskites[11], special attention has been paid to the driving force provided by the local electric field since it is closely relevant to the practical device operations. This local electric field can be generated by the excited-state charge carriers trapped in the defect sites of mixed-halide perovskites, and it is capable of promoting the migration of halide vacancies and anions to trigger the phase segregation process[12]. For mixed-halide perovskites, the defect sites could be located on the surfaces[13] or in the grain boundaries[14] of polycrystalline films, and across the side edges[15] or within the internal volumes[16] of single crystals. The lack of a general consensus on the exact locations of the defect sites impedes further explorations of their dynamic interactions with the charge carriers, whose transition from the trapping to de-trapping statuses determines the survival time of a local electric field driving the phase segregation process.

When the size of mixed-halide perovskites is reduced in all three dimensions to the Bohr diameter scale, the as-formed nanocrystals (NCs) are rendered an additional knob of quantum confinement to tune the emission wavelength in addition to the compositional change[17].

Moreover, the increased surface-to-volume ratio in mixed-halide perovskite NCs leads to a significant dependence of their optoelectronic properties on the external surfaces, which have been well documented in the traditional CdSe NCs to hold abundant defect sites for the effective trapping of the excited-state charge carriers[18-20]. The local electric field created as such in a single CdSe NC is capable of triggering the spectral diffusion effect, which is manifested as a stochastic photoluminescence (PL) peak shift due to the appearance, movement and disappearance of the surface-trapped charge carriers[21-23]. Since their external surfaces are the dominant sources of defect sites to capture the excited-state charge carriers, mixed-halide perovskite NCs can provide a much-simplified platform to investigate how the phase segregation process is affected by the local electric fields.

In this work, we focus on the phase segregation studies of mixed-halide perovskite $CsPbBr_{1.2}I_{1.8}$ NCs at room temperature, showing that they can be converted completely to the $CsPbBr_3$ NCs under sufficient time of the laser excitation. Interestingly, when the laser excitation is removed in the middle of the phase segregation process, the bromide-rich phase with a blue-shifted PL peak can still be formed in the dark for as long as several minutes. Without the complicating factor of the excited-state charge carriers inside the $CsPbBr_{1.2}I_{1.8}$ NCs, this prolonged phase segregation should be driven by the local electric fields created by the long-lived charge carriers trapped in the defect sites. After being deposited onto a passivating flake of hexagonal boron nitride (hBN), the $CsPbBr_{1.2}I_{1.8}$ NCs demonstrate suppressed light-induced and dark-prolonged phase segregations processes, confirming that the defect sites capable of trapping photogenerated charge carriers are mainly located on their external surfaces.

**Results**

According to a standard hot-injection method[24] (see experimental methods in the Supplementary Information), the cuboid $CsPbBr_{1.2}I_{1.8}$ NCs are synthesized with an average edge length of ~23 nm (see the transmission electron microscopy image in Supplementary Fig. 1). As shown in Supplementary Fig. 2 for the $CsPbBr_{1.2}I_{1.8}$ NCs contained in a hexane solution, their band-edge absorption and emission peaks are located at ~608 nm and ~634 nm, respectively, with a Stokes shift of ~26 nm that is consistent with those values reported previously in the literature[25]. One drop of this concentrated solution is spin-coated onto a fused silica substrate to form a solid film of the $CsPbBr_{1.2}I_{1.8}$ NCs, which are excited by a 405 nm or 568 nm continuous-wave (CW) laser with the spot size of ~500 nm to study their optical properties at room temperature (see experimental methods in the Supplementary Information).

As can be seen from the time-dependent spectral image plotted in Fig. 1a, the PL peak of the $CsPbBr_{1.2}I_{1.8}$ NCs shifts completely from the initial ~634 nm to the final ~510 nm upon continuous 405 nm laser excitation at the power density of ~50 $W/cm^2$. This implies that each single $CsPbBr_{1.2}I_{1.8}$ NC within the laser spot has finished the phase segregation process, which is signified as a blue shift in the PL peak[26-28] instead of the red shift commonly observed in the bulk counterpart[4]. The fully-segregated $CsPbBr_{1.2}I_{1.8}$ NCs are then left in the dark, with the 405 nm laser being unblocked only for 1 s at several time points to record the PL spectral evolution still at the power density of ~50 $W/cm^2$. As shown in Fig. 1b for the phase remixing process, the PL peak shifts continuously to the red side and arrives at ~600 nm after 70 min of dark treatment. Of special note is that in all of the previous optical measurements like the one performed in Fig. 1b, the phase remixing process of mixed-halide perovskites is exclusively monitored after a full completion of the phase segregation process.

To provide more information on the phase remixing process, we now excite the $CsPbBr_{1.2}I_{1.8}$ NCs for only 35 s using the 405 nm laser with a power density of ~50 $W/cm^2$ to induce the partial phase segregation on purpose. The PL peak of the CsPbBr$_{1.2}$I$_{1.8}$ NCs shifts from ~634 nm to ~617 nm in Fig. 2a, the latter of which is far from the ~510 nm wavelength measured in Fig. 1a after the complete phase segregation. In the next, the partially-segregated CsPbBr$_{1.2}$I$_{1.8}$ NCs are left in the dark, while the 405 nm laser is unblocked for only 1 s at several time points to monitor the PL spectral evolution at a power density of ~3 W/cm$^2$, which is not high enough to reach the required threshold for the occurrence of phase segregation[29] (see Supplementary Fig. 3). As can be seen in Fig. 2b, the PL peak continues its blue shift to stop at ~609 nm after 2 min, and then reverts to the red side to reach ~616 nm and ~624 nm at the time points of 3 min and 5 min, respectively. Since there is no laser excitation on the CsPbBr$_{1.2}$I$_{1.8}$ NCs, there must exist some residual forces to drive their dark phase segregation process without the influence of the excited-state charge carriers.

To further corroborate the above point, we alternatively excite the CsPbBr$_{1.2}$I$_{1.8}$ NCs with a 568 nm CW laser, whose photon energy is smaller than the bandgap energy of the CsPbBr$_3$ NCs formed after the complete phase segregation. The power density of the 568 nm laser is set at a higher value of ~5000 W/cm$^2$, which can promote the same rate of phase segregation as that associated with the 405 nm laser at ~50 W/cm$^2$. In the course of continuous excitation of the CsPbBr$_{1.2}$I$_{1.8}$ NCs, the 568 nm laser is blocked for only 1 s at several time points to allow the PL spectral measurements with the 405 nm laser at a power density of ~3 W/cm$^2$. As shown in Fig. 2c (see Supplementary Fig. 4 for another example), the PL peak shifts from ~634 nm to ~561 nm after 15 s of phase segregation, the latter of which has a photon energy just a little larger than that of the 568 nm laser. Beyond this time point, the PL peak continues its blue shift to arrive at ~533 nm and ~525 nm after 2 min and 4 min of the 568 nm laser excitation, respectively. Since no thermal effect is caused to the CsPbBr$_{1.2}$I$_{1.8}$ NCs under below-bandgap laser excitation (see Supplementary Fig. 5), we can conclude again that their prolonged blue shifts of the PL peak must be driven by some other factors without photogeneration of the excited-state charge carriers. These $CsPbBr_{1.2}I_{1.8}$ NCs are then left in the dark, with the 405 nm laser being unblocked for only 1 s at several time points to acquire the PL spectra with a power density of ~3 W/cm$^2$. As can be seen in Fig. 2d, the PL peak shifts a little bit further to the blue side to stop at ~520 nm after 2 min, and then reverts to the red side to arrive at ~527 nm, ~557 nm and ~571 nm after 3 min, 12 min and 26 min, respectively.

Since phase segregation of the $CsPbBr_{1.2}I_{1.8}$ NCs can still proceed in the dark or under below-bandgap laser excitation, the experimental results demonstrated in Fig. 2 can be utilized to exclude most of the underlying mechanisms proposed so far for mixed-halide perovskites. These mainly include the polaronic strain field[30,31], the iodide oxidation or repulsion[32,33] and even the system free-energy variation[34,35], whose functionalities are all dependent on the very existence of the excited-state charge carriers. As such, the local electric field induced by the trapped charge carriers is left as the main candidate to support the prolonged phase segregation in the dark (see the schematic diagrams in Fig. 3). Upon laser excitation of a single $CsPbBr_{1.2}I_{1.8}$ NC, some of the excited-state charge carriers move to its surface and get trapped by the defect sites[21-23] (Fig. 3a). The local electric fields created therein are capable of breaking the Pb-I bonds[26], with the freed iodide anions migrating to the surroundings of this single NC[28] (Fig. 3b). After the laser excitation has been removed, the number of the surface-trapped charge carriers decreases slowly over time, while the residual electric fields can still break the Pb-I bonds to release more iodide anions (Fig. 3c). After the surface-trapped charge carriers have disappeared completely, the iodide anions will start filling the internal NC vacancies under the entropic driving force to start the phase remixing process (Fig. 3d). As has been observed in other low-dimensional semiconductor nanostructures[36], it would take several to tens of minutes for their surface charges to get totally dissipated by means of thermal fluctuations[37] and scattering interactions with the surrounding air[38]. Moreover, the time duration of several minutes measured here for the prolonged phase segregation is comparable to the dwelling time of the surface-trapped charge carriers estimated previously for the single CdSe NCs[39].

For semiconductor perovskite materials, it is well known that the application of an electric field can cause the lattice distortion effect[40,41]. Meanwhile, the strain field in mixed-halide perovskites is frequently invoked as the driving force of phase segregation[28,30,31,42]. Upon the removal of laser excitation on the $CsPbBr_{1.2}I_{1.8}$ NCs, it is possible for the surface-trapped charge carriers to disappear instantly together with the induced local electric fields. In this case, the partially-segregated $CsPbBr_{1.2}I_{1.8}$ NCs can still undergo further phase segregation in the dark, under the assumption that it may take some time for the lattice distortions and thus the associated strain fields to be fully relaxed. To rule out this possibility, we deposit a solid film of the $CsPbBr_{1.2}I_{1.8}$ NCs between two electrodes biased at the electric field of ~2 V/μm to induce phase segregation in the dark[26] (see experimental methods in the Supplementary Information). By exciting the $CsPbBr_{1.2}I_{1.8}$ NCs for only 1 s at several time points with the 405 nm laser at a power density of ~3 W/cm$^2$, we can see from the obtained PL spectra in Fig. 4a that the PL peak has shifted from ~634 nm to ~617 nm after 2 min of the electrical biasing operation. Unlike the blue shift observed in Fig. 2b upon the removal of laser excitation, the PL peak of the partially-segregated $CsPbBr_{1.2}I_{1.8}$ NCs now demonstrates a prompt red shift in Fig. 4b after the electric field has been turned off, moving from ~617 nm to ~634 nm within 2 min of the phase remixing process. We can then safely conclude that there are no residual lattice distortions or strain fields to maintain the prolonged phase segregation in the dark, which should be contributed solely by the local electric fields created by the long-lived surface-trapped charge carriers.

**Discussion**

In the end, we would like to elaborate a little more on the defect sites in the $CsPbBr_{1.2}I_{1.8}$ NCs, which are proposed in the traditional CdSe NCs to be mainly on their external surfaces[18-23,39]. In a recent work with high-resolution structural characterizations, the plane defects were discovered within the internal volume of a single $CsPbBr_{1.2}I_{1.8}$ NC to divide it into multiple emissive regions[43]. As shown in Supplementary Fig. 6a for a single $CsPbBr_{1.2}I_{1.8}$ NC excited at ~3 W/cm$^2$ by a 405 nm pulsed laser, the $g^{(2)}(0)$ value is estimated to be ~0.636 from the second-order photon correlation measurement (see experimental methods in the Supplementary Information). This low purity of single-photon emission (~36.4%) is in stark contrast to that of ~97.8% measured in Supplementary Fig. 6b for a single $CsPbBr_3$ NC under the same experimental conditions, implying that there do exist multiple emissive regions and thus the plane defects inside a single $CsPbBr_{1.2}I_{1.8}$ NC. Consequently, it is unclear whether the excited-state charge carriers are trapped by the defect sites on the external surface or within the internal volume of a single $CsPbBr_{1.2}I_{1.8}$ NC.

To clarify the above point, we insert an hBN flake (see Supplementary Fig. 7 for the optical microscope image) between the fused silica substrate and a solid film of the $CsPbBr_{1.2}I_{1.8}$ NCs. When excited by the 405 nm CW laser at a power density of ~50 W/cm$^2$, these $CsPbBr_{1.2}I_{1.8}$ NCs suffer from a much weaker phase segregation effect (Fig. 4c) than those deposited directly on top of the fused silica substrate (Fig. 4d). Moreover, no further phase segregation is observed in the partially-segregated $CsPbBr_{1.2}I_{1.8}$ NCs above the hBN flake once the excitation laser has been blocked. The suppressed phase segregation under both laser excitation and dark treatment suggests that the defect sites of the $CsPbBr_{1.2}I_{1.8}$ NCs should be mostly on their external surfaces, which can be effectively passivated by the hBN flake[44,45] to reduce the number of trapped charge carriers and the strength of the local electric fields.

To summarize, we have induced partial instead of complete phase segregation in mixed-halide perovskite $CsPbBr_{1.2}I_{1.8}$ NCs, observing that they can still undergo the halide demixing process for several minutes after being left in the dark. Besides the $CsPbBr_{1.2}I_{1.8}$ NCs focused here, this prolonged phase segregation is also possessed by the $CsPbBr_xI_{3-x}$ NCs with other x values (see Supplementary Fig. 8). Since there are no excited-state charge carriers inside these $CsPbBr_xI_{3-x}$ NCs, the dark phase segregation should be sustained by the local electric fields from the surface-trapped charge carriers, resulting in the breaking of Pb-I bonds and the generation of more freed iodide anions. The above conclusion can be naturally extended to the laser excitation case, pinpointing local electric field as the only underlying force to drive phase segregation of the $CsPbBr_xI_{3-x}$ NCs. From our optical measurements on the mixed-halide $CsPbBr_{1.5}I_{1.5}$ microplates, the phase remixing process starts immediately upon the removal of laser excitation (see Supplementary Fig. 9). This suggests that the local electric fields may have limited access to the whole bulk volume within the laser spot size of ~500 nm, while they are sufficient enough to drive the halide migration process in a single $CsPbBr_xI_{3-x}$ NC with the average edge length of ~23 nm. Consequently, the phase segregation in bulk mixed-halide perovskites could be jointly contributed by various underlying mechanisms, among which the influence of local electric field has been well understood in the current work to greatly mitigate the future research efforts.

**Data availability**

The data supporting the findings of this study are available from the corresponding authors upon request.

**References**

1. Lai, M., Shin, D., Jibril, L. & Mirkin, C. A. Combinatorial synthesis and screening of mixed halide perovskite megalibraries. *J. Am. Chem. Soc.* **144,** 13823-13830 (2022).

2. Nedelcu, G., Protesescu, L., Yakunin, S., Bodnarchuk, M. I., Grotevent, M. J. & Kovalenko, M. V. Fast anion-exchange in highly luminescent nanocrystals of cesium lead halide perovskites ($CsPbX_3$, X= Cl, Br, I). *Nano Lett.* **15,** 5635-5640 (2015).

3. Xing, G., Mathews, N., Lim, S. S., Yantara, N., Liu, X., Sabba, D., Grätzel, M., Mhaisalkar, S. & Sum, T. C. Low-temperature solution-processed wavelength-tunable perovskites for lasing. *Nat. Mater.* **13,** 476-480 (2014).

4. Hoke, E. T., Slotcavage, D. J., Dohner, E. R., Bowring, A. R., Karunadasa, H. I. & McGehee, M. D. Reversible photo-induced trap formation in mixed-halide hybrid perovskites for photovoltaics. *Chem. Sci.* **6,** 613-617 (2015).

5. Braly, I. L., Stoddard, R. J., Rajagopal, A., Uhl, A. R., Katahara, J. K., Jen, A. K.-Y. & Hillhouse, H. W. Current-induced phase segregation in mixed halide hybrid perovskites and its impact on two-terminal tandem solar cell design. *ACS Energy Lett.* **2,** 1841-1847 (2017).

6. Al-Ashouri, A., Köhnen, E., Li, B., Magomedov, A., Hempel, H., Caprioglio, P., Márquez, J. A., Vilches, A. B. M., Kasparavicius, E., Smith, J. A., Phung, N., Menzel, D., Grischek, M., Kegelmann, L., Skroblin, D., Gollwitzer, C., Malinauskas, T., Jošt, M., Matic, G., Rech, B., Schlatmann, R., Topic, M., Korte, L., Abate, A., Stannowski, B., Neher, D., Stolterfoht, M., Unold, T., Getautis, V. & Albrecht, S. Monolithic perovskite/silicon tandem solar cell with >29% efficiency by enhanced hole extraction. *Science* **370,** 1300-1309 (2020).

7. Feldmann, S., Macpherson, S., Senanayak, S. P., Abdi-Jalebi, M., Rivett, J. P., Nan, G., Tainter, G. D., Doherty, T. A. S., Frohna, K., Ringe, E., Friend, R. H., Sirringhaus, H.,

Saliba, M., Beljonne, D., Stranks, S. D. & Deschler, F. Photodoping through local charge carrier accumulation in alloyed hybrid perovskites for highly efficient luminescence. *Nat. Photon.* **14,** 123-128 (2020).

8. Hassan, Y., Park, J. H., Crawford, M. L., Sadhanala, A., Lee, J., Sadighian, J. C., Mosconi, E., Shivanna, R., Radicchi, E., Jeong, M., Yang, C., Choi, H., Park, S. H., Song, M. H., De Angelis, F., Wong, C. Y., Friend, R. H., Lee, B. R. & Snaith, H. J. Ligand-engineered bandgap stability in mixed-halide perovskite LEDs. *Nature* **591,** 72-77 (2021).

9. Surendran, A., Yu, X., Begum, R., Tao, Y., Wang, Q. J. & Leong, W. L. All inorganic mixed halide perovskite nanocrystal-graphene hybrid photodetector: from ultrahigh gain to photostability. *ACS Appl. Mater. Interfaces* **11,** 27064-27072 (2019).

10. Wang, Y., Zhang, X., Wang, D., Li, X., Meng, J., You, J., Yin, Z. & Wu, J. Compositional engineering of mixed-cation lead mixed-halide perovskites for high-performance photodetectors. *ACS Appl. Mater. Interfaces* **11,** 28005-28012 (2019).

11. Brennan, M. C., Ruth, A., Kamat, P. V. & Kuno, M. Photoinduced anion segregation in mixed halide perovskites. *Trends Chem.* **2,** 282-301 (2020).

12. Knight, A. J., Wright, A. D., Patel, J. B., McMeekin, D. P., Snaith, H. J., Johnston, M. B. & Herz, L. M. Electronic traps and phase segregation in lead mixed-halide perovskite. *ACS Energy Lett.* **4,** 75-84 (2019).

13. Belisle, R. A., Bush, K. A., Bertoluzzi, L., Gold-Parker, A., Toney, M. F. & McGehee, M. D. Impact of surfaces on photoinduced halide segregation in mixed-halide perovskites. *ACS Energy Lett.* **3,** 2694-2700 (2018).

14. Tang, X., van den Berg, M., Gu, E., Horneber, A., Matt, G. J., Osvet, A., Meixner, A. J., Zhang, D. & Brabec, C. J. Local observation of phase segregation in mixed-halide perovskite. *Nano Lett.* **18,** 2172-2178 (2018).

15. Chen, W., Mao, W., Bach, U., Jia, B. & Wen, X. Tracking dynamic phase segregation in mixed-halide perovskite single crystals under two-photon scanning laser illumination. *Small Methods* **3,** 1900273 (2019).

16. Mao, W., Hall, C. R., Chesman, A. S. R., Forsyth, C., Cheng, Y.-B., Duffy, N. W., Smith, T. A. & Bach, U. Visualizing phase segregation in mixed-halide perovskite single crystals. *Angew. Chem., Int. Ed.* **58,** 2893-2898 (2019).

17. Protesescu, L., Yakunin, S., Bodnarchuk, M. I., Krieg, F., Caputo, R., Hendon, C. H., Yang, R. X., Walsh, A. & Kovalenko, M. V. Nanocrystals of cesium lead halide perovskites ($CsPbX_3$, X = Cl, Br, and I): novel optoelectronic materials showing bright emission with wide color gamut. *Nano Lett.* **15,** 3692-3696 (2015).

18. Whitham, P. J., Knowles, K. E., Reid, P. J. & Gamelin, D. R. Photoluminescence blinking and reversible electron trapping in copper-doped CdSe nanocrystals. *Nano Lett.* **15,** 4045-4051 (2015).

19. Efros, A. L. & Nesbitt, D. J. Origin and control of blinking in quantum dots. *Nat. Nanotechnol.* **11,** 661-671 (2016).

20. Saniepay, M., Mi, C., Liu, Z., Abel, E. P. & Beaulac, R. Insights into the structural complexity of colloidal CdSe nanocrystal surfaces: correlating the efficiency of nonradiative excited-state processes to specific defects. *J. Am. Chem. Soc.* **140,** 1725-1736 (2018).

21. Empedocles, S. A., Norris, D. J. & Bawendi, M. G. Photoluminescence spectroscopy of single CdSe nanocrystallite quantum dots. *Phys. Rev. Lett.* **77,** 3873-3876 (1996).

22. Hinterding, S. O. M., Salzmann, B. B. V., Vonk, S. J. W., Vanmaekelbergh, D., Weckhuysen, B. M., Hutter, E. M. & Rabouw, F. T. Single trap states in single CdSe nanoplatelets. *ACS Nano* **15,** 7216-7225 (2021).

23. Panfil, Y. E., Cui, J., Koley, S. & Banin, U. Complete mapping of interacting charging states in single coupled colloidal quantum dot molecules. *ACS Nano* **16,** 5566-5576 (2022).

24. Wang, H., Sui, N., Bai, X., Zhang, Y., Rice, Q., Seo, F. J., Zhang, Q., Colvin, V. L. & Yu, W. W. Emission recovery and stability enhancement of inorganic perovskite quantum dots. *J. Phys. Chem. Lett.* **9,** 4166-4173 (2018).

25. Brennan, M. C., Herr, J. E., Nguyen-Beck, T. S., Zinna, J., Draguta, S., Rouvimov, S., Parkhill, J. & Kuno, M. Origin of the size-dependent Stokes shift in $CsPbBr_3$ perovskite nanocrystals. *J. Am. Chem. Soc.* **139,** 12201-12208 (2017).

26. Zhang, H., Fu, X., Tang, Y., Wang, H., Zhang, C., Yu, W. W., Wang, X., Zhang, Y. & Xiao, M. Phase segregation due to ion migration in all-inorganic mixed-halide perovskite nanocrystals. *Nat. Commun.* **10,** 1088 (2019).

27. Wu, D., Li, N., Liu, B., Guan, J., Li, M., Yan, L., Wang, J., Peng, S., Wang, B., Dong, H., Du, X., Guo, S. & Yang, W. Time-resolved study of laser-induced phase separation in $CsPb(I_xBr_{1-x})_3$ perovskite under high pressure. *Appl. Phys. Lett.* **124,** 031109 (2024).

28. Feng, S., Ju, Y., Duan, R., Man, Z., Li, S., Hu, F., Zhang, C., Tao, S., Zhang, W., Xiao, M. & Wang, X. Complete suppression of phase segregation in mixed-halide perovskite nanocrystals under periodic heating. *Adv. Mater.* **36,** 2308032 (2024).

29. Ruth, A., Brennan, M. C., Draguta, S., Morozov, Y. V., Zhukovskyi, M., Janko, B., Zapol, P. & Kuno, M. Vacancy-mediated anion photosegregation kinetics in mixed halide hybrid perovskites: coupled kinetic Monte Carlo and optical measurements. *ACS Energy Lett.* **3,** 2321-2328 (2018).

30. Bischak, C. G., Hetherington, C. L., Wu, H., Aloni, S., Ogletree, D. F., Limmer, D. T. & Ginsberg, N. S. Origin of reversible photoinduced phase separation in hybrid perovskites. *Nano Lett.* **17,** 1028-1033 (2017).

31. Bischak, C. G., Wong, A. B., Lin, E., Limmer, D. T., Yang, P. & Ginsberg, N. S. Tunable polaron distortions control the extent of halide demixing in lead halide perovskites. *J. Phys. Chem. Lett.* **9,** 3998-4005 (2018).

32. Mathew, P. S., Samu, G. F., Janáky, C. & Kamat, P. V. Iodine (I) expulsion at photoirradiated mixed halide perovskite interface. should I stay or should I go? *ACS Energy Lett.* **5,** 1872-1880 (2020).

33. Frolova, L. A., Luchkin, S. Y., Lekina, Y., Gutsev, L. G., Tsarev, S. A., Zhidkov, I. S., Kurmaev, E. Z., Shen, Z. X., Stevenson, K. J., Aldoshin, S. M. & Troshin, P. A. Reversible $Pb^{2+}/Pb^{0}$ and $I^{-}/I^{3-}$ redox chemistry drives the light-induced phase segregation in all-inorganic mixed halide perovskites. *Adv. Energy Mater.* **11,** 2002934 (2021).

34. Draguta, S., Sharia, O., Yoon, S. J., Brennan, M. C., Morozov, Y. V., Manser, J. S., Kamat, P. V., Schneider, W. F. & Kuno, M. Rationalizing the light-induced phase separation of mixed halide organic-inorganic perovskites. *Nat. Commun.* **8,** 200 (2017).

35. Chen, Z., Brocks, G., Tao, S. & Bobbert, P. A. Unified theory for light-induced halide segregation in mixed halide perovskites. *Nat. Commun.* **12,** 2687 (2021).

36. Li, S., Liao, K., Bi, Y., Ding, K., Sun, E., Zhang, C., Wang, L., Hu, F., Xiao, M. & Wang, X. Optical readout of charge carriers stored in a 2D memory cell of monolayer $WSe_2$, *Nanoscale* **16,** 3668-3675 (2024).

37. Wang, Q., Wen, Y., Cai, K., Cheng, R., Yin, L., Zhang, Y., Li, J., Wang, Z., Wang, F., Wang, F., Shifa, T. A., Jiang, C., Yang, H. & He, J. Nonvolatile infrared memory in $MoS_2$/PbS van der Waals heterostructures. *Sci. Adv.* **4,** eaap7916 (2018).

38. Tan, Y., Yu, K., Yang, T., Zhang, Q., Cong, W., Yin, H., Zhang, Z., Chen, Y. & Zhu, Z. The combinations of hollow $MoS_2$ micro@nano-spheres: one-step synthesis, excellent photocatalytic and humidity sensing properties. *J. Mater. Chem. C* **2,** 5422-5430 (2014).

39. Empedocles, S. A. & Bawendi, M. G. Quantum-confined Stark effect in single CdSe nanocrystallite quantum dots. *Science* **278,** 2114-2117 (1997).

40. Chen, B., Li, T., Dong, Q., Mosconi, E., Song, J., Chen, Z., Deng, Y., Liu, Y., Ducharme, S., Gruverman, A., De Angelis, F. & Huang, J. Large electrostrictive response in lead halide perovskites. *Nat. Mater.* **17,** 1020-1026 (2018).

41. Rana, S., Awasthi, K., Bhosale, S. S., Diau, E. W.-G. & Ohta, N. Temperature-dependent electroabsorption and electrophotoluminescence and exciton binding energy in $MAPbBr_3$ perovskite quantum dots. *J. Phys. Chem. C* **123,** 19927-19937 (2019).

42. Mao, W., Hall, C. R., Bernardi, S., Cheng, Y.-B., Widmer-Cooper, A., Smith, T. A. & Bach, U. Light-induced reversal of ion segregation in mixed-halide perovskites. *Nat. Mater.* **20,** 55-61 (2021).

43. Liu, J., Zhu, C., Poles, M., Zhang, Z., Wang, L., Zhang, C., Liu, Z., Tao, S., Xiao M. & Wang, X. Discrete elemental distributions inside a single mixed-halide perovskite nanocrystal for the self-assembly of multiple quantum-light sources. *Nano Lett.* **23,** 10089-10096 (2023).

44. Lee, G.-H., Yu, Y.-J., Cui, X., Petrone, N., Lee, C.-H., Choi, M. S., Lee, D.-Y., Lee, C., Yoo, W. J., Watanabe, K., Taniguchi, T., Kim, P. & Hone, J. Flexible and transparent $MoS_2$ field-effect transistors on hexagonal boron nitride-graphene heterostructures. *ACS Nano* **7,** 7931-7936 (2013).

45. Canet-Albiach, R., Krecmarova, M., Bailach, J. B., Gualdrón-Reyes, A. F., Rodríguez-Romero, J., Gorji, S., Pashaei-Adl, H., Mora-Seró, I., Martinez Pastor, J. P., Sánchez-Royo, J. F. & Muñoz-Matutano, G. Revealing giant exciton fine-structure splitting in two-dimensional perovskites using van der Waals passivation. *Nano Lett.* **22,** 7621-7627 (2022).

**Acknowledgements**

This work is supported by the National Basic Research Program of China (Nos. 2019YFA0308704 and 2021YFA1400803), the National Natural Science Foundation of China (Nos. 62174081 and 61974058), and the Priority Academic Program Development of Jiangsu Higher Education Institutions.

**Author contributions**

X.W., C.Z., K.X. and F.H. conceived and designed the experiments. X.M., Y.Y., S.F. and Y.X. prepared the samples and performed the optical measurements. X.M., Y.Y. and X.W. analyzed the data. X.W., X.M., Y.Y. and M.X. co-wrote the manuscript.

**Competing Interests**

The authors declare no competing financial or non-financial interests.

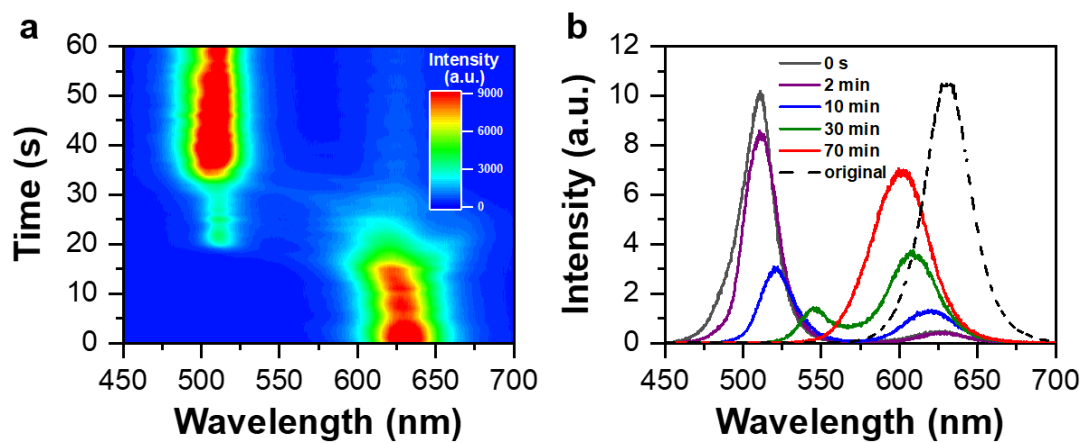

**Figure 1 | Fundamental phase segregation and remixing properties. a,** Time-dependent evolution of the 60 PL spectra each acquired with an integration time of 1 s for the CsPbBr$_{1.2}$I$_{1.8}$ NCs. The phase segregation process completes after ~30 s continuous excitation of the CsPbBr$_{1.2}$I$_{1.8}$ NCs by a 405 nm CW laser at the power density of ~50 W/cm$^2$. **b,** PL spectra measured for these CsPbBr$_{1.2}$I$_{1.8}$ NCs after the excitation laser has then been blocked for 0 s, 2 min, 10 min, 30 min and 70 min, respectively. At each of the above time points, the 405 nm laser is unblocked for 1 s to acquire the PL spectrum still at the power density of ~50 W /cm$^2$. For comparison, the original PL spectrum measured at the beginning of laser excitation in **a** is also provided (black dashed line).

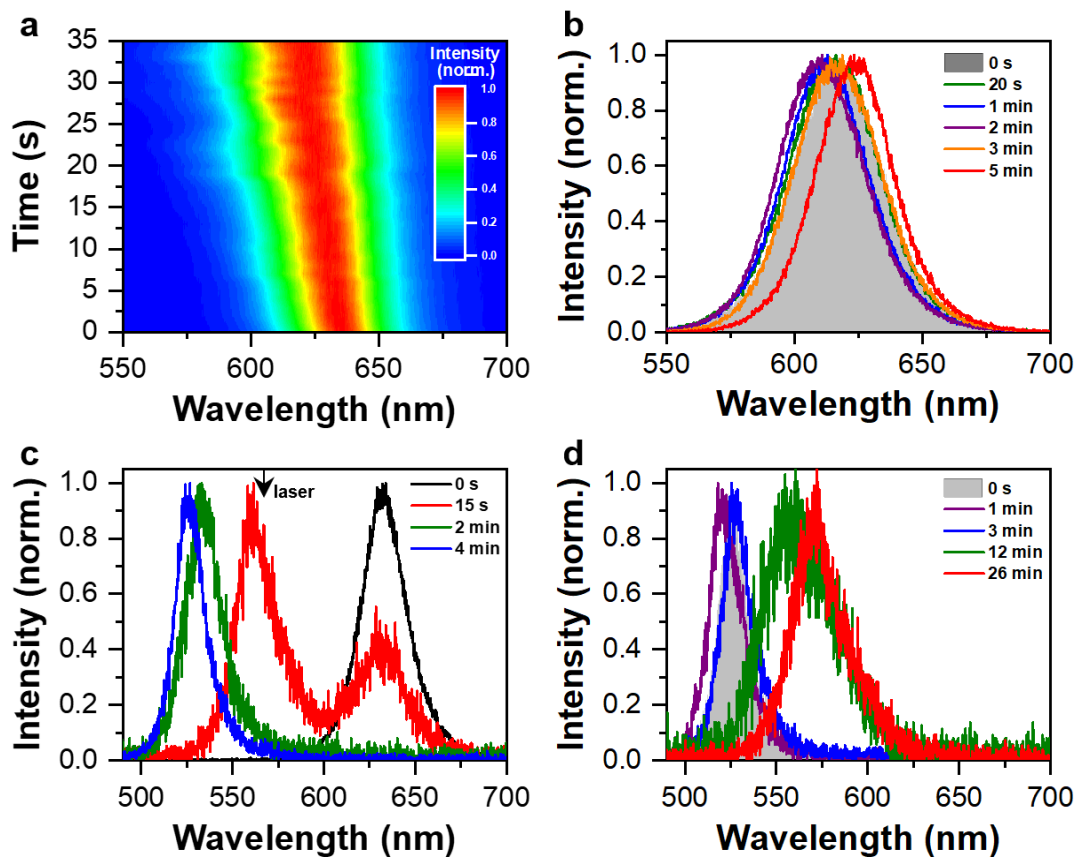

**Figure 2 | Prolonged phase segregation in the dark. a,** Time-dependent evolution of the 35 PL spectra each acquired with an integration time of 1 s for the CsPbBr$_{1.2}$I$_{1.8}$ NCs. The phase segregation process is incomplete after ~30 s continuous excitation of the CsPbBr$_{1.2}$I$_{1.8}$ NCs by a 405 nm CW laser at the power density of ~50 W/cm$^2$. **b,** PL spectra measured for these CsPbBr$_{1.2}$I$_{1.8}$ NCs after the 405 nm laser has then been blocked for 0 s, 20 s, 1 min, 2 min, 3 min and 5 min, respectively. At each of the above time points, the 405 nm laser is unblocked for 1 s to acquire the PL spectrum at the power density of ~3 W/cm$^2$. **c,** PL spectra measured for the CsPbBr$_{1.2}$I$_{1.8}$ NCs after being excited by a 568 nm CW laser at the power density of ~5000 W/cm$^2$ for 0 s, 15 s, 2 min and 4 min, respectively. At each of the above time points, the 568 nm laser is blocked for 1 s while the 405 nm laser is employed to acquire a PL spectrum at the power density of ~3 W/cm$^2$. The solid black arrow on top marks the wavelength position of the 568 nm excitation laser. **d,** PL spectra measured for these CsPbBr$_{1.2}$I$_{1.8}$ NCs after the 568 nm laser has then been blocked for 0 s, 1 min, 3 min, 12 min and 26 min, respectively. At each of the above time points, the 405 nm CW laser is unblocked for 1 s to acquire a PL spectrum at the power density of ~3 W/cm$^2$.

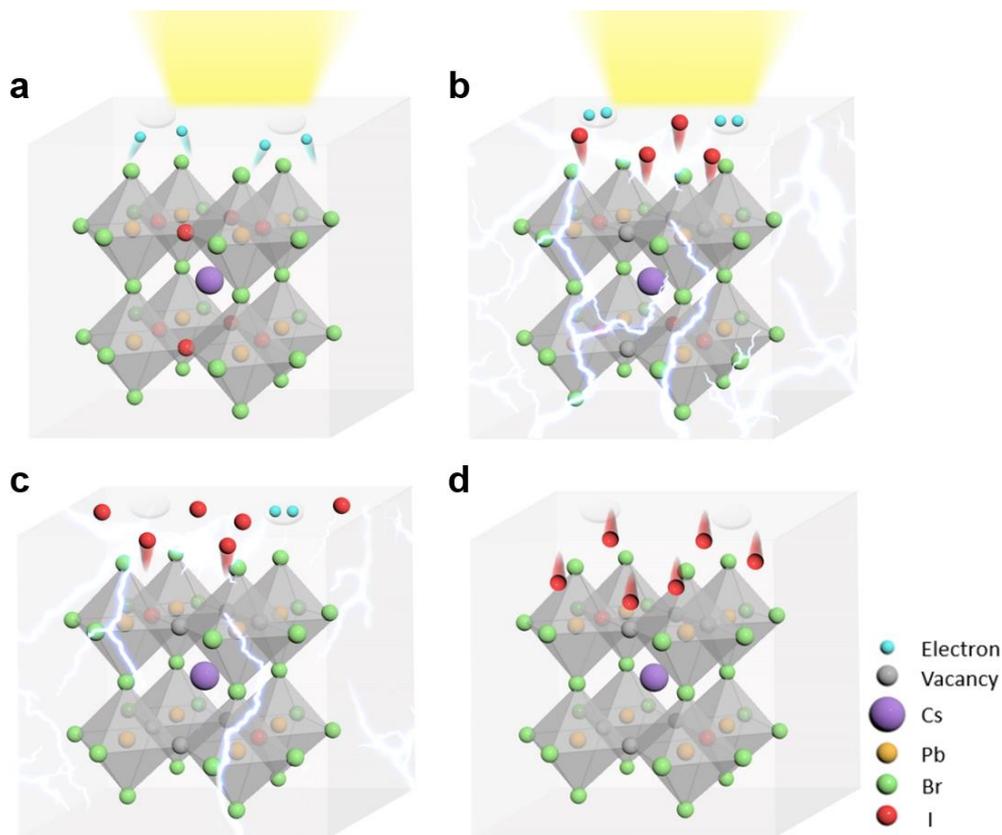

**Figure 3 | Underlying mechanism for the prolonged phase segregation in the dark. a,** Upon laser excitation of a single CsPbBr$_{1.2}$I$_{1.8}$ NC, some of the excited-state charge carriers move to the surface. **b,** The local electric fields created by the charge carriers trapped in the surface defect sites are capable of breaking the Pb-I bonds, with the freed iodide anions migrating to the surroundings of this single NC. **c,** The surface-trapped charge carriers disappear slowly after the removal of laser excitation, with the residual ones still exerting the local electric fields on the single NC to maintain the prolonged phase segregation in the dark. **d,** After the surface-trapped charge carriers totally disappear to terminate the local electric fields, the freed iodide anions move back to the single NC to start the phase remixing process.

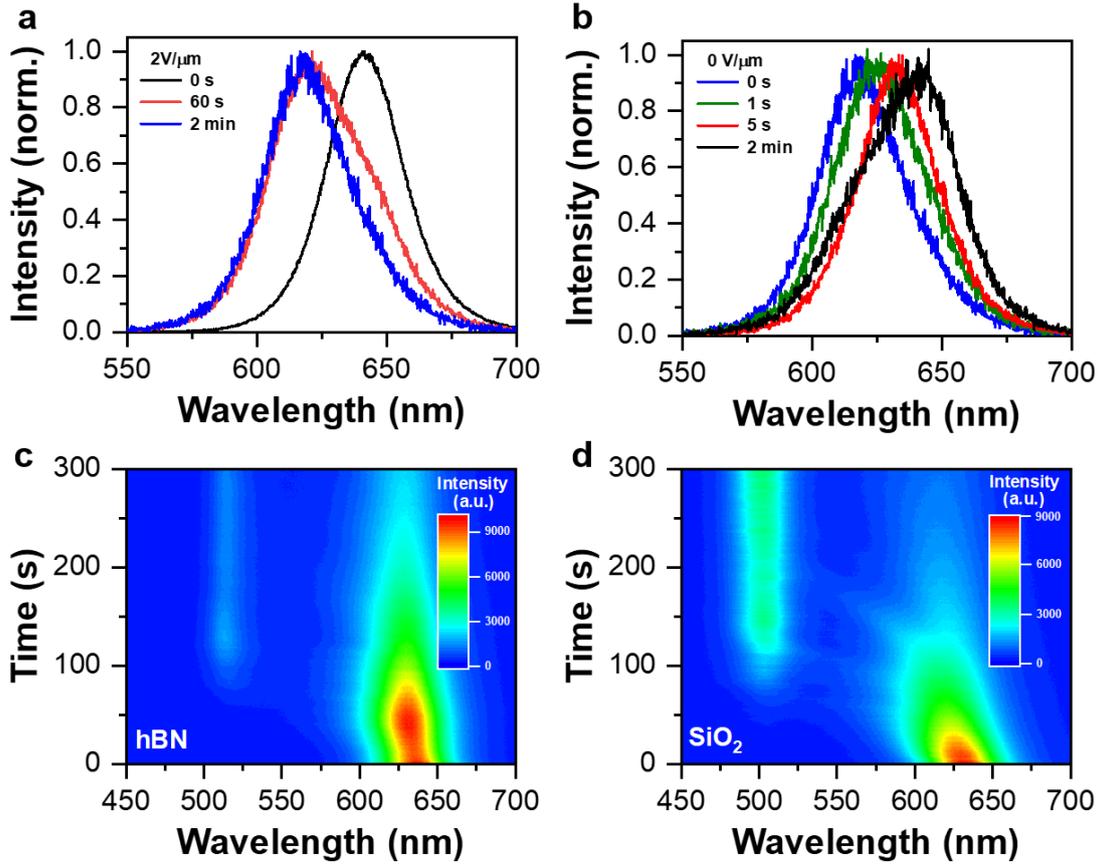

**Figure 4 | Influence of the electric fields on phase segregation. a,** PL spectra measured for the CsPbBr$_{1.2}$I$_{1.8}$ NCs after an electric field of 2 V/μm has been applied for 0 s, 60 s and 2 min in the dark. **b,** PL spectra measured for these CsPbBr$_{1.2}$I$_{1.8}$ NCs after the electric field has then been removed for 0 s, 1 s, 5 s and 2 min in the dark. At each time point in **a** and **b**, a 405 nm CW laser is unblocked for 1 s to acquire the PL spectrum at the power density of ~3 W/cm$^2$. Time-dependent evolutions of the 300 PL spectra each acquired with an integration time of 1 s for the CsPbBr$_{1.2}$I$_{1.8}$ NCs deposited on **c,** the hBN flake and **d,** the SiO$_2$ substrate, respectively. In **c** and **d**, the CsPbBr$_{1.2}$I$_{1.8}$ NCs are continuously excited by a 405 nm CW laser with the power density of ~50 W/cm$^2$.

**Supplementary Information**

**Prolonged Phase Segregation of Mixed-Halide Perovskite Nanocrystals in the Dark**

Xueying Ma[1,†], Yuhui Ye[1,†], Yang Xiao[1], Shengnan Feng[1], Chunfeng Zhang[1], Keyu Xia[2], Fengrui Hu[2], Min Xiao[1,3,*], and Xiaoyong Wang[1,*]

[1]*National Laboratory of Solid State Microstructures, School of Physics, and Collaborative Innovation Center of Advanced Microstructures, Nanjing University, Nanjing 210093, China*

[2]*College of Engineering and Applied Sciences, Nanjing University, Nanjing 210093, China*

[3]*Department of Physics, University of Arkansas, Fayetteville, Arkansas 72701, USA*

[*]Correspondence to M.X. (mxiao@uark.edu) or X.W. (wxiaoyong@nju.edu.cn)

[†]These authors contributed equally to this work

**EXPERIMENTAL METHODS**

**Chemical synthesis.** 0.814 g $Cs_2CO_3$, 2.5 mL oleic acid (OA) and 40 mL octadecene (ODE) were loaded into a 100 mL 3-neck flask, which was degassed for 10 min to evaporate the contained moisture. After switching between the vacuum and $N_2$ environments for 3-5 times, the above solution was heated at 120 °C for 1 h to obtain the Cs-oleate precursor in vacuum. In the next, 0.105 g $PbI_2$, 0.052 g $PbBr_2$, 1.0 mL OA, 1.0 mL oleylamine (OAm) and 10 mL ODE were loaded into a 50 mL 3-neck flask, which was also degassed for 10 min to evaporate the contained moisture. After switching between the vacuum and $N_2$ environments for 3-5 times, this mixture was first heated at 120 °C for 1 h in vacuum and then at 160 °C for 10 min in $N_2$. Right after this operation, 1.0 mL of the Cs-oleate precursor preheated to 120 °C was quickly injected into the above solution and the reaction lasted for 5 s before being stopped by an ice bath. The obtained product was centrifuged for 20 min at 5000 rpm, with the precipitates being dispersed in 5 mL hexane and centrifuged again for 20 min at 5000 rpm. Finally, the supernatant was extracted to get the $CsPbBr_{1.2}I_{1.8}$ NCs mainly used in the experiment. To synthesize the $CsPbBr_{0.9}I_{2.1}$/$CsPbBr_{1.5}I_{1.5}$ NCs, quite similar procedure was followed except that the amounts of $PbI_2$ and $PbBr_2$ were changed to 0.119 g/0.085 g and 0.041 g/0.067 g, respectively.

**Optical measurement.** One drop of the concentrated solution of mixed-halide perovskite NCs, with the composition of $CsPbBr_{1.2}I_{1.8}$, $CsPbBr_{0.9}I_{2.1}$ or $CsPbBr_{1.5}I_{1.5}$, was spin-coated onto a fused silica substrate to form a solid film for the optical studies at room temperature. Alternatively, these mixed-halide perovskite NCs could be spin-coated either onto a hBN flake placed above the fused silica substrate by means of the dry transfer method, or between two electrodes thermally evaporated above the fused silica substrate with a separation distance of 5 μm. The sample substrate was attached to a home-built confocal optical microscope, where mixed-halide perovskite NCs were excited by either a 405 nm or a 568 nm CW laser. After passing through an immersion-oil objective with the numerical aperture of ~1.4, the laser beam was focused onto the sample substrate with a spot size of ~500 nm. Optical emission from mixed-halide perovskite NCs was collected by the same objective and sent through a 0.5 m spectrometer to a charge-coupled device (CCD) camera for the PL spectral measurement. For the optical measurements of single $CsPbBr_{1.2}I_{1.8}$ or $CsPbBr_3$ NCs using a 405 nm pulsed laser with the repetition rate of 5 MHz, one drop of their diluted solution was spin-coated onto a fused silica substrate to form a low-density solid film. The single perovskite NC was studied in the confocal optical microscope with quite similar optical setups to those describe above, except that its optical emission was sent directly to two avalanche photodetectors (APDs) for the second-order photon correlation measurement with a time resolution of ~100 ps.

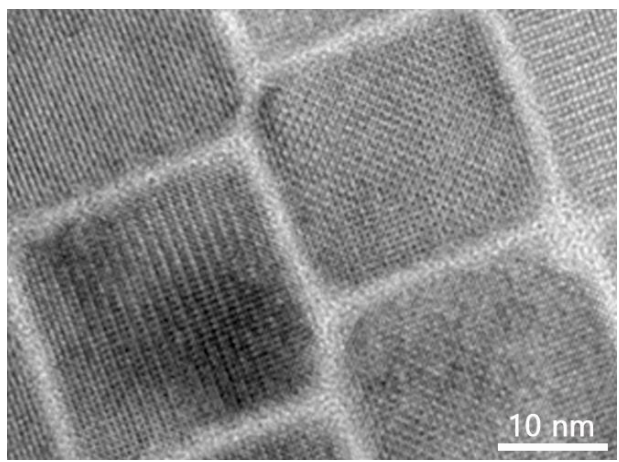

**Supplementary Fig. 1.** Transmission electron microscope image measured for the cuboid CsPbBr$_{1.2}$I$_{1.8}$ NCs with an average edge length of ~23 nm.

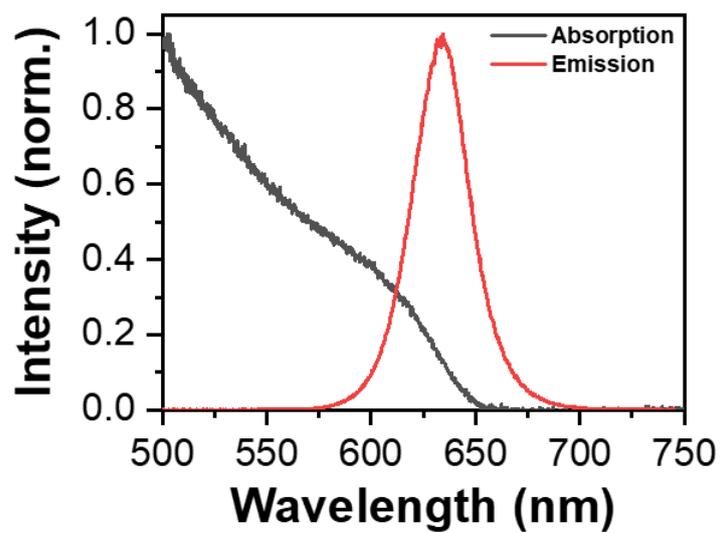

**Supplementary Fig. 2.** Absorption and emission spectra measured for the $CsPbBr_{1.2}I_{1.8}$ NCs in a hexane solution.

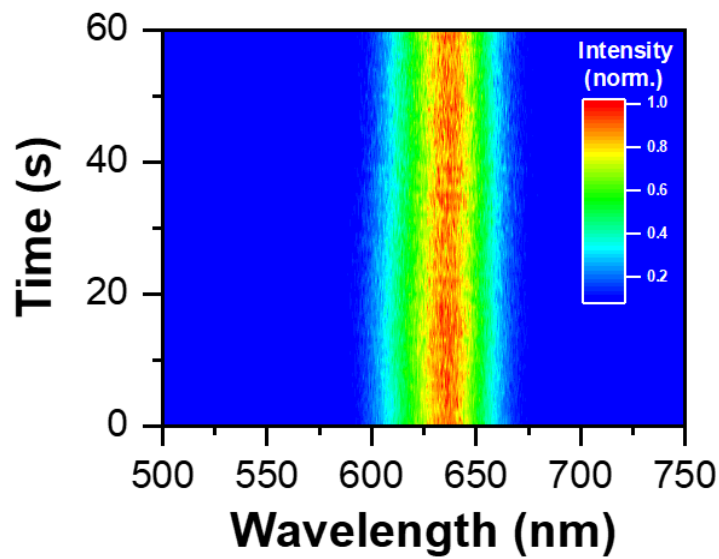

**Supplementary Fig. 3.** Time-dependent evolution of the 60 PL spectra each acquired with an integration time of 1 s for the CsPbBr$_{1.2}$I$_{1.8}$ NCs. These CsPbBr$_{1.2}$I$_{1.8}$ NCs show no sign of phase segregation under the continuous excitation of a 405 nm CW laser at the power density of ~3 W/cm$^2$.

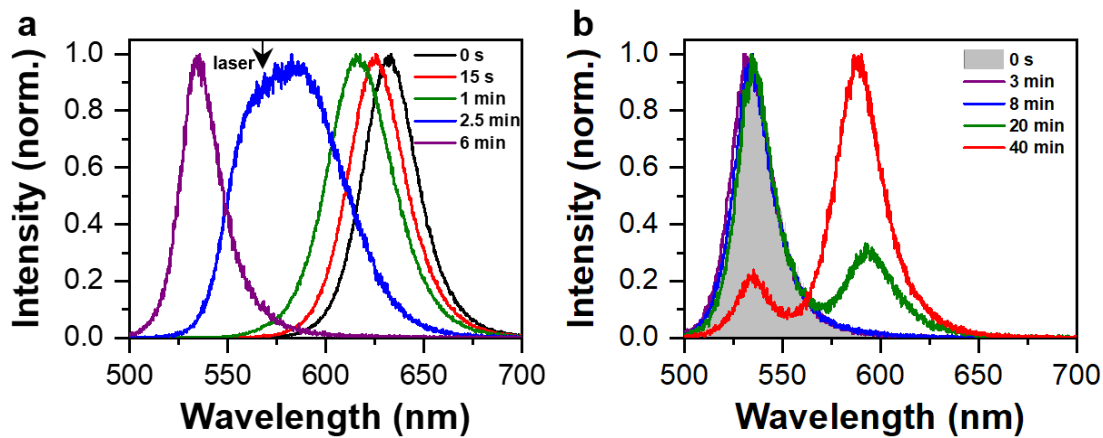

**Supplementary Fig. 4. a,** PL spectra measured for the CsPbBr$_{1.2}$I$_{1.8}$ NCs after being excited by a 568 nm CW laser at the power density of ~5000 W/cm$^2$ for 0 s, 15 s, 1 min, 2.5 min and 6 min, respectively. At each of the above time points, the 568 nm laser is blocked for 1 s while a 405 nm CW laser is employed to acquire the PL spectrum at the power density of ~3 W /cm$^2$. The solid black arrow on top marks the wavelength position of the 568 nm excitation laser. **b,** PL spectra measured for these CsPbBr$_{1.2}$I$_{1.8}$ NCs after the 568 nm laser has then been blocked for 0 s, 3 min, 8 min, 20 min and 40 min, respectively. At each of the above time points, a 405 nm CW laser is unblocked for 1 s to acquire the PL spectrum at the power density of ~3 W /cm$^2$.

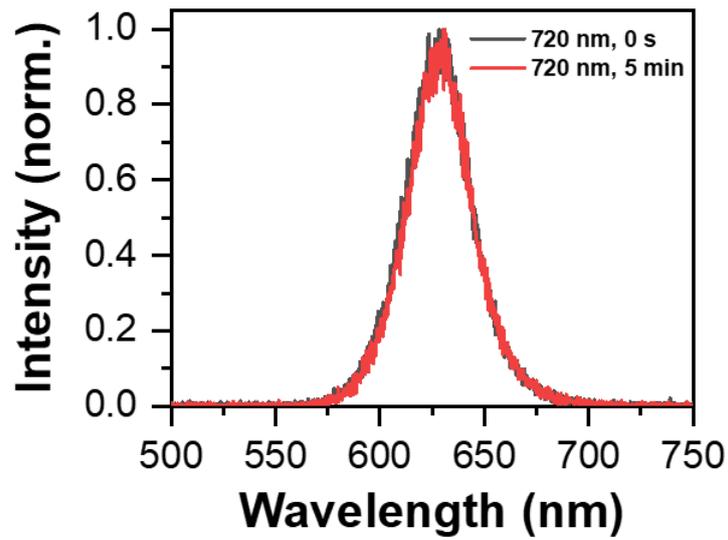

**Supplementary Fig. 5.** PL spectra measured for the CsPbBr$_{1.2}$I$_{1.8}$ NCs after being excited for 0 s and 5 min by a below-bandgap 720 nm pulsed laser at the power density of ~5000 W/cm$^2$, showing that there is no thermal effect to induced the phase segregation process. At each of these two time points, a 405 nm CW laser is unblocked for 1 s to acquire the PL spectrum at the power density of ~3 W /cm$^2$.

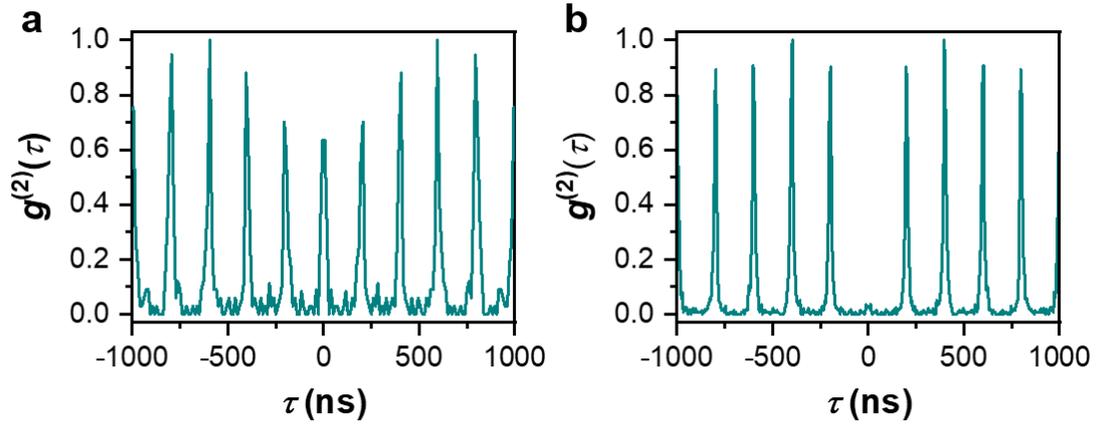

**Supplementary Fig. 6. a,** Second-order photon correlation measurement of a single CsPbBr$_{1.2}$I$_{1.8}$ NC with the $g^{(2)}(0)$ value of ~0.636 and the single-photon emission purity of ~36.4%. **b,** Second-order photon correlation measurement of a single CsPbBr$_3$ NC with the $g^{(2)}(0)$ value of ~0.022 and the single-photon emission purity of ~97.8%. The single CsPbBr$_{1.2}$I$_{1.8}$ or CsPbBr$_3$ NC is excited at room temperature by a 5 MHz 405 nm pulsed laser at the power density of ~3 W /cm$^2$.

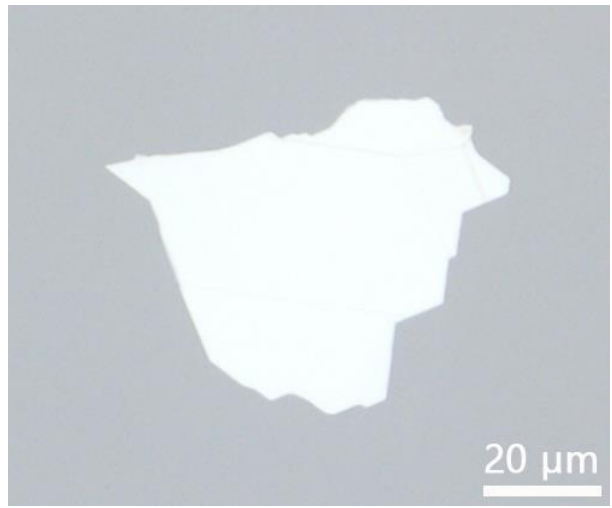

**Supplementary Fig. 7.** Optical microscope image of an hBN flake (white color) with the thickness of ~150 nm.

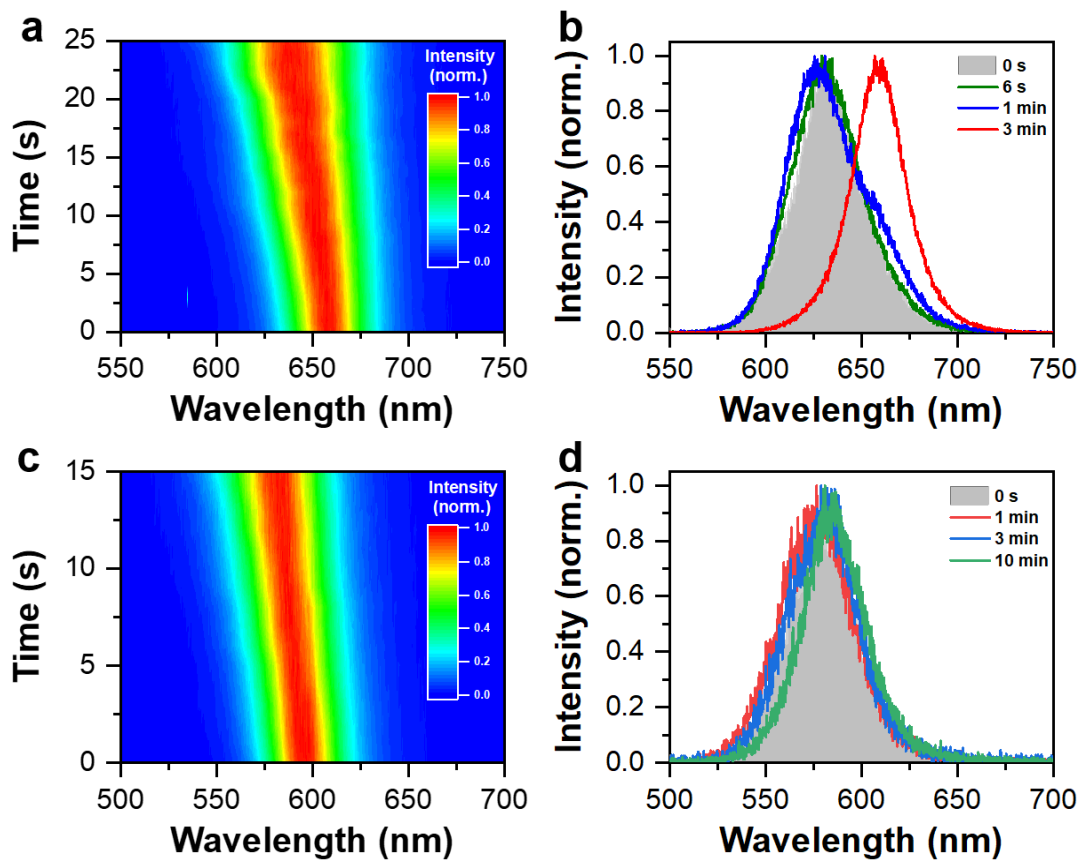

**Supplementary Fig. 8. a,** Time-dependent evolution of the 25 PL spectra each acquired with an integration time of 1 s for the $CsPbBr_{0.9}I_{2.1}$ NCs. The phase segregation process is incomplete after 25 s continuous excitation of the $CsPbBr_{0.9}I_{2.1}$ NCs by a 405 nm CW laser at the power density of ~50 W/cm$^2$. **b,** PL spectra measured for these $CsPbBr_{0.9}I_{2.1}$ NCs after the 405 nm laser has then been blocked for 0 s, 6 s, 1 min and 3 min, respectively. At each of the above time points, the 405 nm laser is unblocked for 1 s to acquire the PL spectrum at a lower power density of ~3 W/cm$^2$. **c,** Time-dependent evolution of the 15 PL spectra each acquired with an integration time of 1 s for the $CsPbBr_{1.5}I_{1.5}$ NCs. The phase segregation process is incomplete after 15 s continuous excitation of the $CsPbBr_{1.5}I_{1.5}$ NCs by a 405 nm CW laser at the power density of ~50 W/cm$^2$. **d,** PL spectra measured for these $CsPbBr_{1.5}I_{1.5}$ NCs after the 405 nm laser has then been blocked for 0 s, 1 min, 3 min and 10 min, respectively. At each of the above time points, the 405 nm laser is unblocked for 1 s to acquire the PL spectrum at a lower power density of ~3 W/cm$^2$.

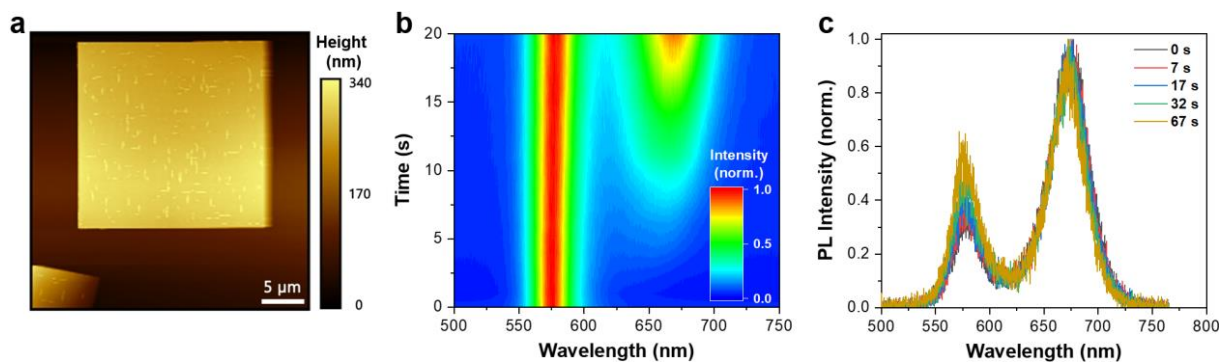

**Supplementary Fig. 9. a,** Atomic force microscope image measured for an individual $CsPbBr_{1.5}I_{1.5}$ microplate. **b,** Time-dependent evolution of the 20 PL spectra each acquired with an integration time of 1 s for this microplate. Manifested as the appearance of a red-shifted PL peak, the phase segregation process is incomplete after this microplate has been continuously excited for 20 s by a 405 nm CW laser at the power density of ~50 W/cm$^2$. **c,** PL spectra measured for the partially-segregated microplate after the 405 nm laser has then been blocked for 0 s, 7 s, 17 s, 32 s and 67 s, respectively. At each of the above time points, the 405 nm laser is unblocked for 1 s to acquire the PL spectrum at a lower power density of ~5 W/cm$^2$. Normalized with respect to the red-shifted PL peak, these PL spectra demonstrate that the PL intensity of the original $CsPbBr_{1.5}I_{1.5}$ composition is instantly recovered upon removal of the laser excitation.

12